\pdfoutput=1
\documentclass[12pt]{iopart}

\usepackage{graphicx,xcolor}
\expandafter\let\csname equation*\endcsname\relax
\expandafter\let\csname endequation*\endcsname\relax
\usepackage{amsmath, amsthm, amssymb}
\UseRawInputEncoding

\begin{document}
	
	\title{Effects of zero and reversed magnetic shear on resistive wall modes in a limiter tokamak plasma}
	
	\author{Sui Wan}
	
	\address{State Key Laboratory of Advanced Electromagnetic Technology, \\International Joint Research Laboratory of Magnetic Confinement Fusion and Plasma Physics, School of Electrical and Electronic Engineering,
		\\	Huazhong University of Science and Technology, Wuhan, 430074,
		China}
	
	\author{Ping Zhu*}
	
	\address{State Key Laboratory of Advanced Electromagnetic Technology, \\International Joint Research Laboratory of Magnetic Confinement Fusion and Plasma Physics, School of Electrical and Electronic Engineering,
		\\	Huazhong University of Science and Technology, Wuhan, 430074,
		China;
		~\\
		Department of Nuclear Engineering and Engineering Physics, 
		\\University of Wisconsin-Madison, Madison,
		Wisconsin, 53706, United States of America}
	\ead{zhup@hust.edu.cn}
	
	\author{Haolong Li}
	
	\address{College of Physics and Optoelectronic Engineering, Shenzhen University,
		Shenzhen, 518060,  China}
			
		\author{Rui Han}
		
		\address{CAS Key Laboratory of Geospace Environment and Department of Engineering and Applied Physics, University of Science and Technology of China, Hefei, 230026, China}
	
	\vspace{10pt}
	\begin{indented}
		\item[]January 11th, 2024
	\end{indented}
	\clearpage
	\begin{abstract}
		
		Advanced tokamak scenarios often feature equilibriums with zero and reversed magnetic shear. To isolate and investigate their impacts on the resistive wall mode (RWM) instability analytically, we construct a series of cylindrical limiter equilibriums with reversed magnetic shear in the core and zero magnetic shear towards plasma edge, as a prototype of the configurations in advanced tokamak scenarios. Uniform plasma pressure is assumed, so that we can focus our analysis on the current-driven RWMs. Based on the reduced ideal MHD equations, analytical solutions for the $ n=1 $ resistive wall mode are obtained, which indicate that increasing the reversal of magnetic shear in the core region enhances the RWM instability, whereas the widened region of zero shear near edge leads to lower growth rate of RWM, except when the $ q $ value with zero magnetic shear approaches rational values. On the other hand, enhanced positive shear at plasma edge is found to be stabilizing on RWM. NIMROD calculation results confirm these analytical findings.
	\end{abstract}
	
	%
	\vspace{2pc}
	\noindent{\it Keywords}: RWM, NIMROD, reversed magnetic shear, zero magnetic shear
	%
	
	\submitto{\PPCF}
	%
	%
	%
	\section{Introduction}
	
	The external kink mode in a cylinder or a torus can be entirely suppressed when a perfect conducting wall is located sufficiently close to the plasma edge. However, due to the finite resistivity of a less perfect conducting wall, the residual external kink mode interacts with the conductor wall with a slowed-down growth rate and is referred to as the resistive wall mode (RWM), which has a growth rate $ \gamma\sim 1/\tau_{\rm w} $~\cite{joffrin07a,chu10a} that is determined by many factors, including the wall¡¯s position and resistivity. In practice, RWM instability often sets the operation limits and leads to disruptions if left uncontrolled.
	
	The advanced tokamak (AT) scenario stands as the favored approach for future tokamaks, including the International Thermonuclear Experimental Reactor (ITER)    and the China Fusion Engineering Test Reactor (CFETR). The primary objective of the AT scenario is to optimize the current distribution with the increased share of bootstrap current along with the reduced demand for auxiliary current drives, which in turn, creates a reversed magnetic shear configuration. However, owing to the highly peaked pressure gradient and a substantial off-axis bootstrap current, AT scenarios are more susceptible to the effects of RWMs~\cite{hender07a}. In addition, the existence of a region with weak magnetic shear at the plasma edge pedestal may give rise to a pressure driven, infernal component of the external mode, which can significantly destabilize the RWM, particularly in the flat-$q$ region near rational surfaces, leading to a lower $\beta$ limit~\cite{zheng13a, zheng13b, han23a}. Analyzing the stability of RWMs in presence of zero and reversed magnetic shear for the advanced tokamak scenarios is thus crucial to the accurate assessment of the stable operation regime of future tokamak based fusion reactors.

	To study the above effects on RWM, we consider a cylindrical limiter tokamak plasma equilibrium with a circular cross-section. The simple analytical equilibrium has a reversed magnetic shear in the core region and a zero or weak shear near the edge, which is meant to serve as a proto-type of the toroidal configuration in the AT scenarios, so that the weak and reversed shear effects can be amenable to analytical calculations. We compare the analytical dispersion relations and the numerical calculations developed for the low-$ \beta $ and current-driven RWMs in such a cylindrical model configuration. These results may provide some insights to other and future studies on the RWM behavior in a divertor tokamak plasma.

	The rest of this paper is organized as follows. Section~\ref{sec:2} and \ref{sec:3} introduce the equilibrium and analytical dispersion relations for RWM with different safety factor profiles in cylindrical plasma with zero $ \beta $ based on Finn's model~\cite{finn95a}. Section~\ref{sec:4} describes the numerical calculation method using the NIMROD code~\cite{sovinec04a}. Section~\ref{sec:5} presents the analytical and numerical results, as well as their comparison. Finally, Section~\ref{sec:6} provides the conclusions and discussion on the findings.
	
	\section{Equilibrium}\label{sec:2}
	
\begin{figure}
		\centering
		\includegraphics[width=0.7\linewidth]{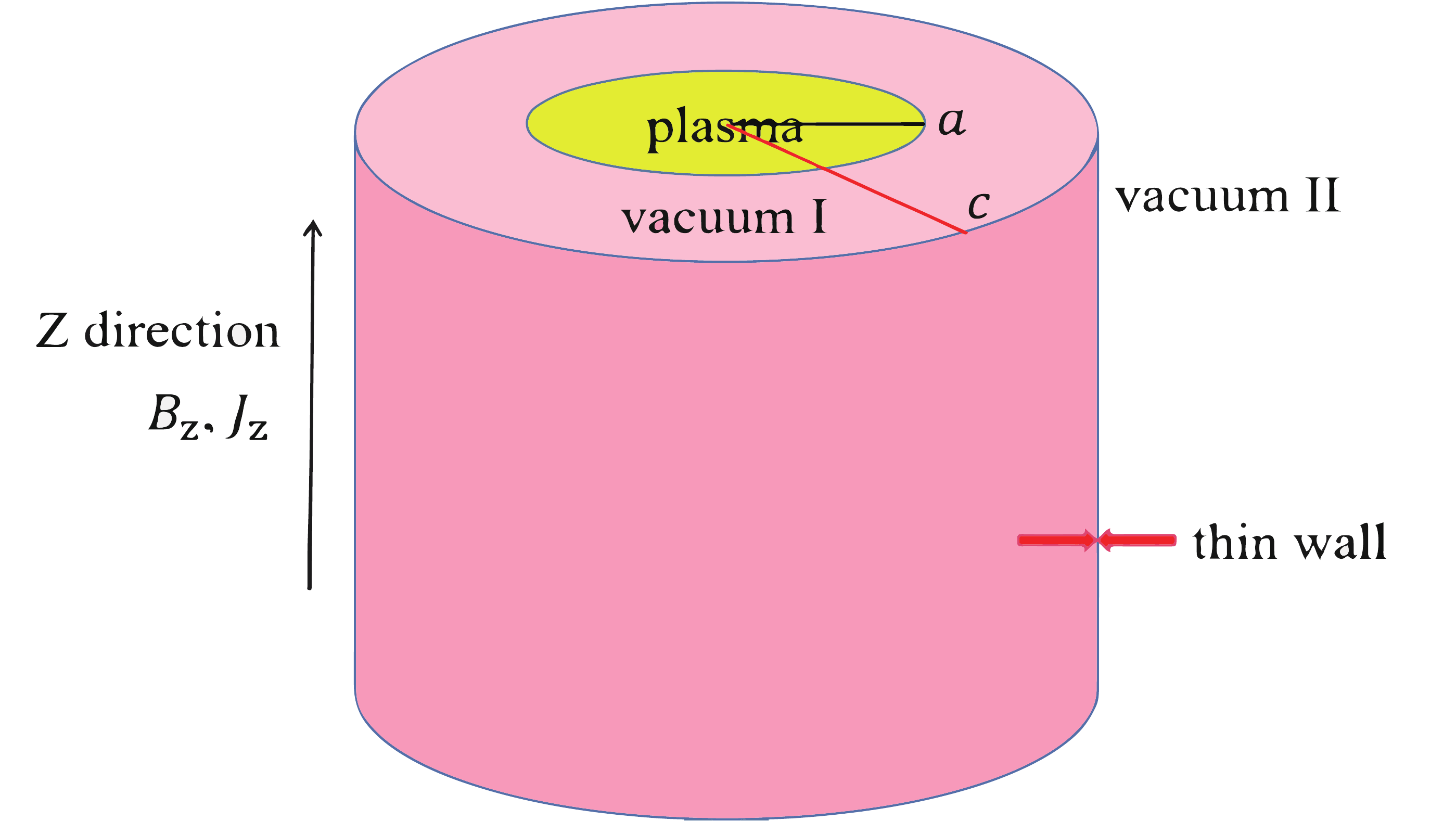}
		\caption{Schematic illustration of cylindrical plasma configuration used in analytical and computational model.}
		\label{fig:configuration}
	\end{figure}

	We consider a cylindrical plasma with a minor radius $ a $ and a periodical length $ 2\pi R $ surrounded by a vacuum region and a thin conducting wall located at radius $ r_w $, where $ R $  denotes the major radius (figure~\ref{fig:configuration}). The cylindrical coordinate system ($ r, \theta, z $) is adopted here. The equilibrium satisfies
	
	\begin{equation}\label{eq:force-balance1}
		\frac{\rm d}{{\rm d} r}\left(p+\frac{B_{\theta}^{2}+B_{z}^{2}}{2 \mu_{0}}\right)+\frac{B_{\theta}^{2}}{\mu_{0} r}=0
	\end{equation}
	where $ p $ is the pressure, $ B_{\theta} $ and $ B_z $ are the poloidal and the axial components of magnetic field respectively, and $ \mu_0 $ is the vacuum permeability.

\begin{figure}
		\centering
		\includegraphics[width=0.8\linewidth]{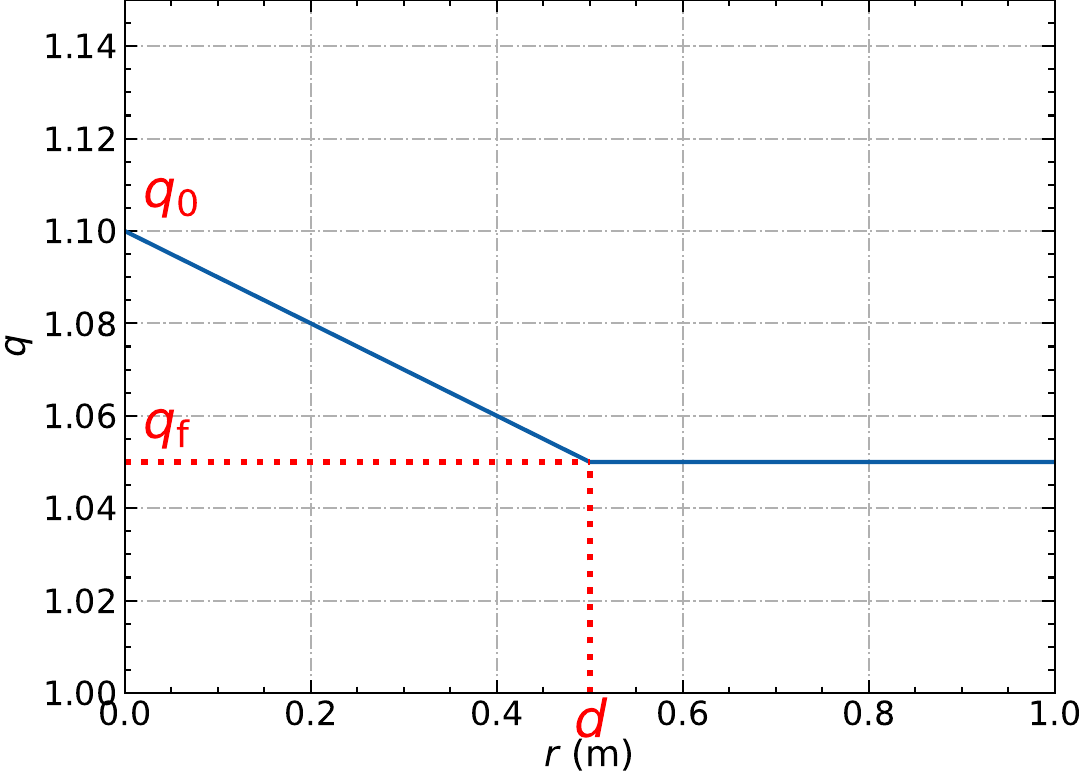}
		\caption{The first type of safety factor profile as a function of the minor radius, where $ q_0=1.1 $, $ q_f=1.05 $, $ d=0.5 $~m.}
		\label{fig:function1}
	\end{figure}

	Since we mainly consider modes driven by the plasma current only, here we assume a uniform pressure profile, i.e, $ \nabla p=0 $ in the plasma region. The safety factor in cylindrical plasma is given by $ q\left( r\right) =rB_z/RB_{\theta} $, which in this study is specified in a piece-wise linear function of radius as (figure~\ref{fig:function1})
	\begin{equation}\label{eq:zhexian representation}
		q(r)= \begin{cases}\frac{q_f-q_0}{d}r+q_0 &0\leq r \leq d \\ q_f & d<r \leq a \end{cases}
	\end{equation}
	where $ q_0 $ and $ q_f $ are the safety factors at the magnetic axis and of the zero magnetic shear region at plasma edge respectively, and $ d $ represents the radial width of the shear reversal region.
	
\begin{figure}
		\centering
		\includegraphics[width=0.8\linewidth]{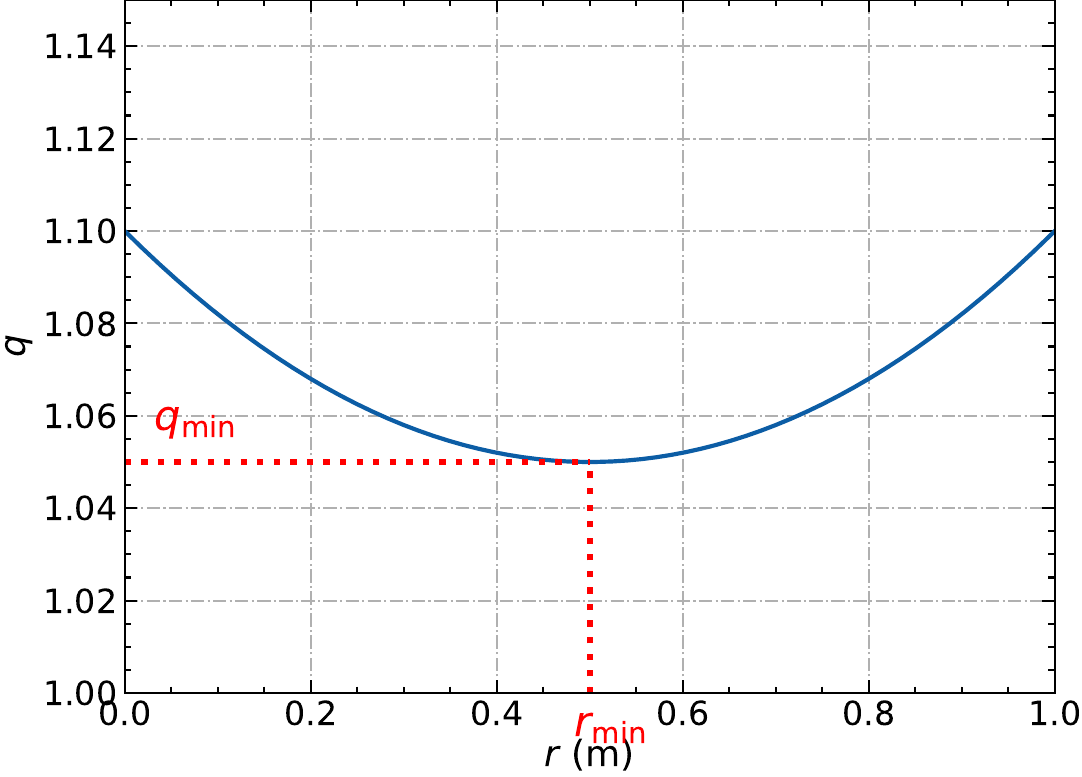}
		\caption{The second type of safety factor profile as a function of the minor radius, where $ a_1=b_2=0.2 $, $ b_1=c_2=-0.2 $, $ c_1=d_2=1.1 $, $ a_2=0 $. }
		\label{fig:function2}
	\end{figure}

	For comparison another set of $q$-profile in the form of higher order polynomials are also considered in analysis (figure~\ref{fig:function2})

	\begin{equation}\label{eq:dxs representation}
		q(r)= \begin{cases}a_1r^2+b_1r+c_1 & 0\leq r \leq r_{\rm min} \\ a_2r^3+b_2r^2+c_2r+d_2 & r_{\rm min}<r \leq a \end{cases}
	\end{equation}
	where $ r_{\rm min} $ refers to the radius of the minimum safety factor.
	\section{Analytical dispersion relations}\label{sec:3}
	
	We consider the perturbation equation for a low-$ \beta $ or force-free equilibrium~\cite{finn95a}
	\begin{equation}\label{eq:disturbance}
		-\frac{\mu_0\gamma^2}{F\left( r\right) }\nabla_{\perp}\cdot\left(\rho\nabla_{\perp}\frac{\delta \psi}{F\left( r\right) } \right) =\nabla_{\perp}^2 \delta \psi-\frac{m\mu_0}{rF\left( r\right) }\frac{\mathrm{d}J_z}{\mathrm{d}r}\delta \psi
	\end{equation}
	where $ J_z $ is the equilibrium plasma current density along the cylinder axis, $ \gamma $ is the growth rate of the instability, $ \delta \psi $ is the perturbed poloidal flux function, which is assumed proportional to $ \exp (im\theta+ikz) $, where $ k=-n/R $, $ m $ and $ n $ are the poloidal and toroidal mode numbers, respectively, and $F(r)=\frac{B_\theta(r)}{r}\left(m-nq(r) \right) $. The opertor $\nabla_{\perp}=-\vec{b}\times\left( \vec{b}\times\nabla\right) $, and $ \nabla^2_{\perp}\delta\psi\equiv \frac{1}{r}\left(r\delta\psi^{\prime}\right)^{\prime}-\frac{m^2}{r^2}\delta\psi -k^2 \delta \psi $.
	
	\subsection{$ q $ profiles in piece-wise linear functions of radius}
	
	First, we consider the analytical $n=1$ RWM dispersion relation for the safety factor profile in the form of a piecewise liner function equation~(\ref{eq:zhexian representation}) and the resistive wall located at $ r_w=c $ which is significantly thinner than the skin depth.
	
	The $ q $ profile in equation~(\ref{eq:zhexian representation}) assumes the presence of a reversed magnetic shear in the core region. When the magnetic shear reversal in the core region is small, the term $\frac{m\mu_0}{rF\left( r\right) }\frac{\mathrm{d}J_z}{\mathrm{d}r} $ can be approximated as a constant $ P $ in the region $ 0\leq r \leq d $ and $ T $ in the region $ d<r \leq a $, $ \mu_0 \rho_0 \gamma^2\sim 0 $ for resistive wall mode and $ k\rightarrow 0 $ in the limit of large aspect ratio. Therefore, equation~(\ref{eq:disturbance}) can be simplified in the plasma region as
	\begin{equation}\label{eq:disturbance-1}
		r^2 \delta \psi^{\prime\prime}+r\delta\psi^{\prime}+\left[\left(-P \right)r^2-m^2  \right]\delta\psi=0, \quad 0\leq r \leq d
	\end{equation}
	
	\begin{equation}\label{eq:disturbance-2}
		r^2 \delta \psi^{\prime\prime}+r\delta\psi^{\prime}+\left[\left(-T \right)r^2-m^2  \right]\delta\psi=0, \quad d<r \leq a
	\end{equation}
	
	The solution for the above equation of the perturbed flux function inside plasma can be approximated as
	\begin{equation}\label{eq:RWM-pert}
		\delta \psi(r)= 
		\begin{cases} \psi_0 {\rm J}_m\left(pr \right) & 0\leq r \leq d \\ \psi_1 {\rm J}_m\left(tr \right)+\psi_2 {\rm Y}_m\left(tr\right) & d< r \leq a 
		\end{cases}
	\end{equation}
	where $ {\rm J}_m $ and $ {\rm Y}_m $ are Bessel functions of the first and the second kinds, respectively,  and $ p=\sqrt{-P}$, $ t=\sqrt{-T}$. The coefficients $ \psi_0 $, $ \psi_1 $ and $ \psi_2 $ are determined using the boundary conditions. In the vacuum region outside plasma, $ \rho=0 $ and $ J_z=0 $, thus equation~(\ref{eq:disturbance}) reduces to 
	\begin{equation}\label{eq:IWM-pert2}
		\nabla^2_{\perp}\delta \psi=0
	\end{equation}
	
	Finally, we obtain the expressions for the perturbed magnetic flux function in different region as follows
	\begin{equation}\label{eq:zx-per}
		\delta \psi(r)= \begin{cases} \psi_0 {\rm J}_m\left(pr \right) & 0\leq r \leq d \\ \psi_1 {\rm J}_m\left(tr \right)+\psi_2 {\rm Y}_m\left(tr\right) & d< r \leq a \\ \psi_3 r^m+\psi_4 r^{-m} & a<r\leq c \\ \psi_5 r^{-m} & r>c \end{cases}
	\end{equation}
	In this case, the perturbed magnetic field penetrate the wall, and its temporal change can induce an electric field as well as an eddy current within the wall~\cite{fitzpatrick96a}
	\begin{equation}\label{eq:RWM-jump1}
		\frac{\partial(\delta \psi)}{\partial t}=\frac{\eta_{w}}{\mu_{0}} \frac{1}{r} \frac{\partial}{\partial r}\left(r \frac{\partial(\delta \psi)}{\partial r}\right)
	\end{equation}
	Assuming $\delta\psi\sim e^{\gamma t}$ and integrating equation~(\ref{eq:RWM-jump1}) through the wall from $ c_- $ to $ c_+ $, we obtain
	\begin{equation}\label{eq:RWM-jump2}
		\gamma \tau_w=\frac{c}{\delta\psi_c}\left[\delta\psi^{\prime} \right]^{c_+}_{c_-}
	\end{equation}
	Here, $\tau_w$ represents the diffusion time of the perturbed magnetic flux in the resistive wall, and $\delta_w$ represents the wall thickness.
	
	Together with the plasma-vacuum jump condition at $ r=a $, the plasma current density discontinuity at $ r=d $, and the assumption $ \mu_0 \rho_0 \gamma^2\rightarrow 0 $, the dispersion relation of the resistive wall mode can be obtained as follows
	\begin{equation}\label{eq:rwm-dr}
		\gamma \tau_w=\frac{-2m\left({\rm J}_m\left(ta\right)+\frac{\psi_2}{\psi_1}{\rm Y}_m\left(ta\right)-\frac{\psi_4}{\psi_1} \right)\left( \frac{c}{a}\right) ^{2m} }{\left({\rm J}_m\left(ta\right)+\frac{\psi_2}{\psi_1}{\rm Y}_m\left(ta\right)-\frac{\psi_4}{\psi_1} \right)\left( \frac{c}{a}\right) ^{2m}+\frac{\psi_4}{\psi_1}}
	\end{equation}
	where
	\begin{equation}\label{eq:rwm-dr1}
		\frac{\psi_2}{\psi_1}=\frac{ \left[ \frac{m\left( \mu_0J_z(d_+)-\mu_0J_z(d_-)\right) }{dF(d)}+\frac{\frac{{\rm d} {\rm J}_m\left(pr\right)}{{\rm d}r}|_{\rm r=d}}{{\rm J}_m\left(pd\right)}\right] {\rm J}_m\left(td\right)-\frac{{\rm d} {\rm J}_m\left(tr\right)}{{\rm d}r}|_{\rm r=d}}{\frac{{\rm d} {\rm Y}_m\left(tr\right)}{{\rm d}r}|_{\rm r=d}-\left[ \frac{m\left( \mu_0J_z(d_+)-\mu_0J_z(d_-)\right) }{dF(d)}+\frac{\frac{{\rm d} {\rm J}_m\left(pr\right)}{{\rm d}r}|_{\rm r=d}}{{\rm J}_m\left(pd\right)}\right] {\rm Y}_m\left(td\right)}
	\end{equation}
	\begin{equation}\label{eq:rwm-dr2}
		\frac{\psi_4}{\psi_1}=\frac{1}{2}\left[ \left( \frac{\mu_0J_z(a)}{F(a)}+1\right)\left( {\rm J}_m\left(ta\right)+\frac{\psi_2}{\psi_1}{\rm Y}_m\left(ta\right)\right)-\frac{1}{m}\left( \frac{{\rm d} {\rm J}_m\left(tr\right)}{{\rm d}r}|_{\rm r=a}+\frac{\psi_2}{\psi_1}\frac{{\rm d} {\rm Y}_m\left(tr\right)}{{\rm d}r}|_{\rm r=a}\right)\right]    
	\end{equation}
	where $ J_z(a) $ is the current density at plasma edge, and $J_z(d-)$ and $ J_z(d+) $ refer to the plasma current density on both sides of the discontinuity at $r=d$ respectively.
	
	\subsection{$q$ profile in higher order polynomials}
	
	The safety factor profile in equation(\ref{eq:dxs representation}) is of higher order polynomials, which ensure the continuity of the current density profile inside plasma. When the magnetic shear is small, $\frac{m\mu_0}{rF\left(r\right)}\frac{\mathrm{d} J_{\mathrm z}}{\mathrm{d} r}$ can be approximated as a constant $Z$ in the plasma region $0\leq r\leq a$ in equation(\ref{eq:disturbance}). As a result, the perturbed magnetic flux function can be approximated using the Bessel function of the first kind
	
	\begin{equation}\label{eq:RWM-pert-dxs}
		\delta \psi=\psi_{01} {\rm J}_m\left(zr \right) ,\quad 0\leq r <a
	\end{equation}
	where $ z=\sqrt{-Z}$, whereas the solutions for the vacuum regions inside and outside the resistive wall remain the same as in the previous subsection. The dispersion relation for the RWM can be thus derived as
	\begin{equation}\label{eq:RWM-DR-dxs}
		\gamma \tau_w=-\frac{-2mA_1\left(\frac{c}{a} \right) ^{2m}}{A_1\left( \frac{c}{a}\right) ^{2m}+1}
	\end{equation}
	with
	\begin{equation}\label{eq:A1}
		A_1=\frac{1+\left(\frac{a}{m{\rm J}_m(za)}\frac{{\rm d}{\rm J}_m(zr)}{{\rm d}r}|_{r=a}-\frac{\mu_0 J_z(a)}{F(a)} \right) }{1-\left(\frac{a}{m{\rm J}_m(za)}\frac{{\rm d}{\rm J}_m(zr)}{{\rm d}r}|_{r=a}-\frac{\mu_0 J_z(a)}{F(a)} \right)} 
	\end{equation}
	
	The above analytical dispersion relations for RWM are compared with the NIMROD calculation results next for verifications.
	
	\section{Numerical model}\label{sec:4}
	
	In this section, we introduce the MHD model for the plasma, wall and vacuum regions implemented in the NIMROD code for the RWM calculations.
	The MHD equations for the plasma region used in our NIMROD calculations are~\cite{sovinec04a}
	
	\begin{gather}
		\frac{\partial N}{\partial t}+\nabla \cdot(N \mathbf{u})=0\\
		M N\left(\frac{\partial}{\partial t}+\mathbf{u} \cdot \nabla\right) \mathbf{u}=\mathbf{J} \times \mathbf{B}-\nabla p-\nabla \cdot \mathbf{\Pi}\\
		\frac{1}{\gamma-1} N\left(\frac{\partial}{\partial t}+\mathbf{u} \cdot \nabla\right) T=-p \nabla \cdot \mathbf{u}\\
		\frac{\partial \mathbf{B}}{\partial t}=-\nabla \times(\eta \mathbf{J}-\mathbf{u} \times \mathbf{B})\\
		\mu_0 \mathbf{J}=\nabla \times \mathbf{B}\\
		\nabla \cdot \mathbf{B}=0
	\end{gather}
	where $ \mathbf{u} $ is the center-of-mass flow velocity, $N$ the plasma number density, $ M $ the ion mass, $ \mathbf{J} $ the plasma current density, $ \mathbf{B} $ the magnetic field, $ p $ the plasma pressure, $ \eta $ the plasma resistivity, $ \mathbf{\Pi} $ the ion viscosity tensor, $ \gamma $ the adiabatic index, and $ \mu_0 $ the vacuum permeability.

	Outside the plasma, a double-wall and double-vacuum region model has been implemented~\cite{sovinec19a}. The inner resistive wall separates the two vacuum regions, whereas the outer ideal perfect conducting wall serves as a fixed outer boundary of the computation domain (figure~\ref{fig:configuration}). If the outer ideal conducting wall is far away from plasma, its influence on the plasma can be neglected, and the inner resistive wall effects on plasma such as those in a vertical displacement event (VDE)~\cite{bunkers20a} can be solely evaluated. The normal components of magnetic field in the two vacuum regions are connected through the resistive wall using
	\begin{equation}\label{eq:wall-treatment1}
		\frac{\partial \left( \mathbf{B}\cdot\hat{\mathbf{n}}\right) }{\partial t}=-\hat{\mathbf{n}}\cdot\boldsymbol{\nabla}\times\left[v_{\text{wall}}\hat{\mathbf{n}}\times\left(\mathbf{B}_{\text{vac}}-\mathbf{B}_{\text{SOL}} \right)  \right] 
	\end{equation}
	where the characteristic wall resistivity parameter $ v_{\text{wall}}=\eta_w/\mu_0\delta_w$, $ \delta_w$ and $ \eta_{w}$ are the wall thickness and resistivity respectively, $\mathbf{n}$ represents the normal unit vector at the wall surface pointing radially outward, $\mathbf{B}_{\text{SOL}}$ and $\mathbf{B}_{\text{vac}}$ represent the magnetic field in the inner vacuum or SOL (scrape-off layer) region and the external vacuum region on both sides of the wall, respectively.

\begin{figure}
		\centering
		\includegraphics[width=0.9\linewidth]{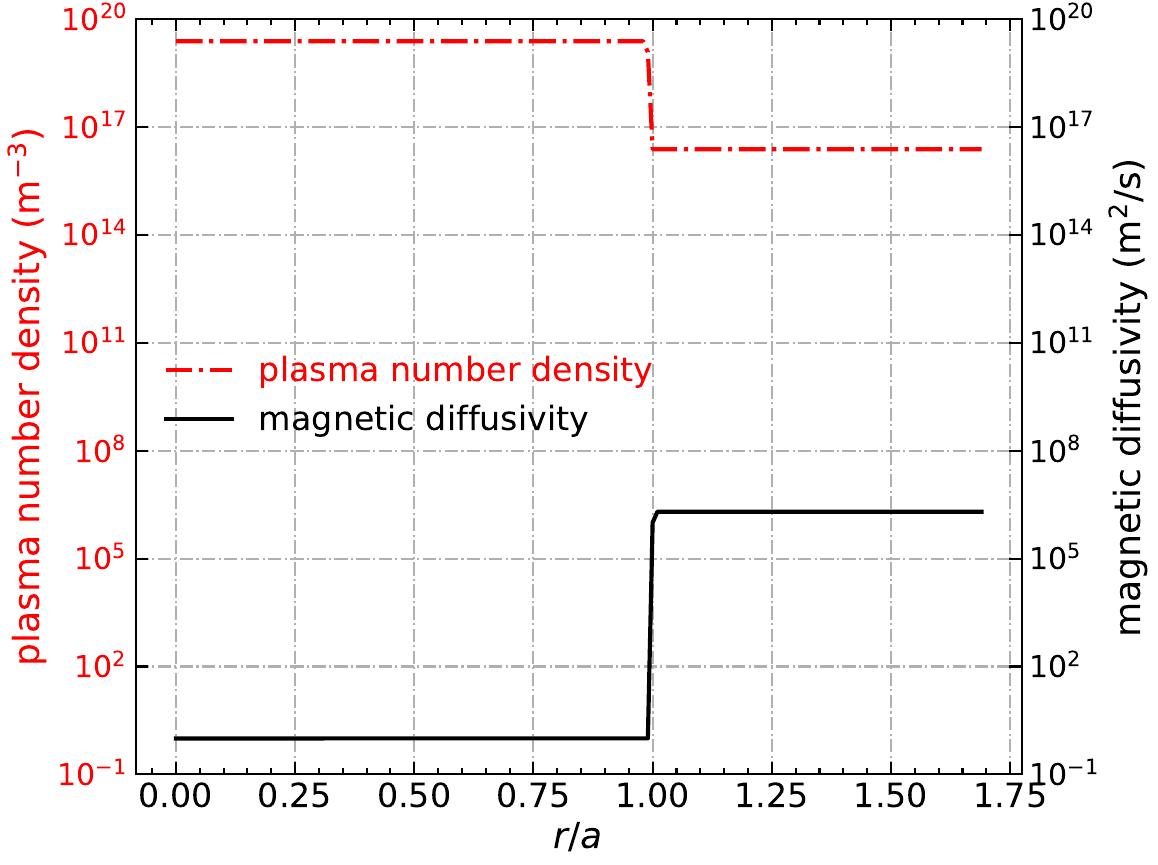}
		\caption{Magnetic diffusivity (black solid line) and number density profiles for plasma and SOL regions as functions of the minor radius (red dotdash line) in NIMROD calculations.}
		\label{fig:density-and-elecd}
	\end{figure}

	The plasma vacuum interface is represented by a steep transition region in the density and resistivity profiles~\cite{han20a, burke10a}. In particular, the resistivity and number density profiles take the following forms
	\begin{gather}\label{eq:profile}
		\eta=\left[ \sqrt{\eta_{\rm plasma}}+\left(\sqrt{\eta_{\rm SOL}}-\sqrt{\eta_{\rm plasma}} \right)r^\alpha \right]^2\\
		n=\left[ \sqrt{n_{\rm plasma}}+\left(\sqrt{n_{\rm SOL}}-\sqrt{n_{\rm plasma}} \right)r^\beta \right]^2
	\end{gather}
	where $ n_{\rm plasma} $, $\eta_{\rm plasma}  $ ($ n_{\rm SOL} $, $\eta_{\rm SOL}  $) are the plasma number density and resistivity in the plasma (SOL) region, respectively. $ \alpha $, $ \beta $ parametrize the gradients of the resistivity and number density profiles at the plasma-vacuum-interface (figure~\ref{fig:density-and-elecd}). The computational results gradually converge with the decreasing plasma resistivity diffusion $ \eta/\mu_0 $ and viscosity within plasma and the increasing resistivity and decreasing plasma number density in the SOL region. In particular, the converged values for $ \eta/\mu_0 $ and viscosity in the plasma region are $ 10^{-5} {\rm m}^2/{\rm s}$ and  $ 10^{-2} {\rm m}^2/{\rm s}$ respectively. The SOL number density needs to reach $ 10^{-4} $ of the plasma density in the core region, and the resistivity diffusion ratio needs to reach $ 10^{12} $ for the calculation results to approach the convergence.

\begin{figure}[h]
		\centering
		\includegraphics[width=1\linewidth]{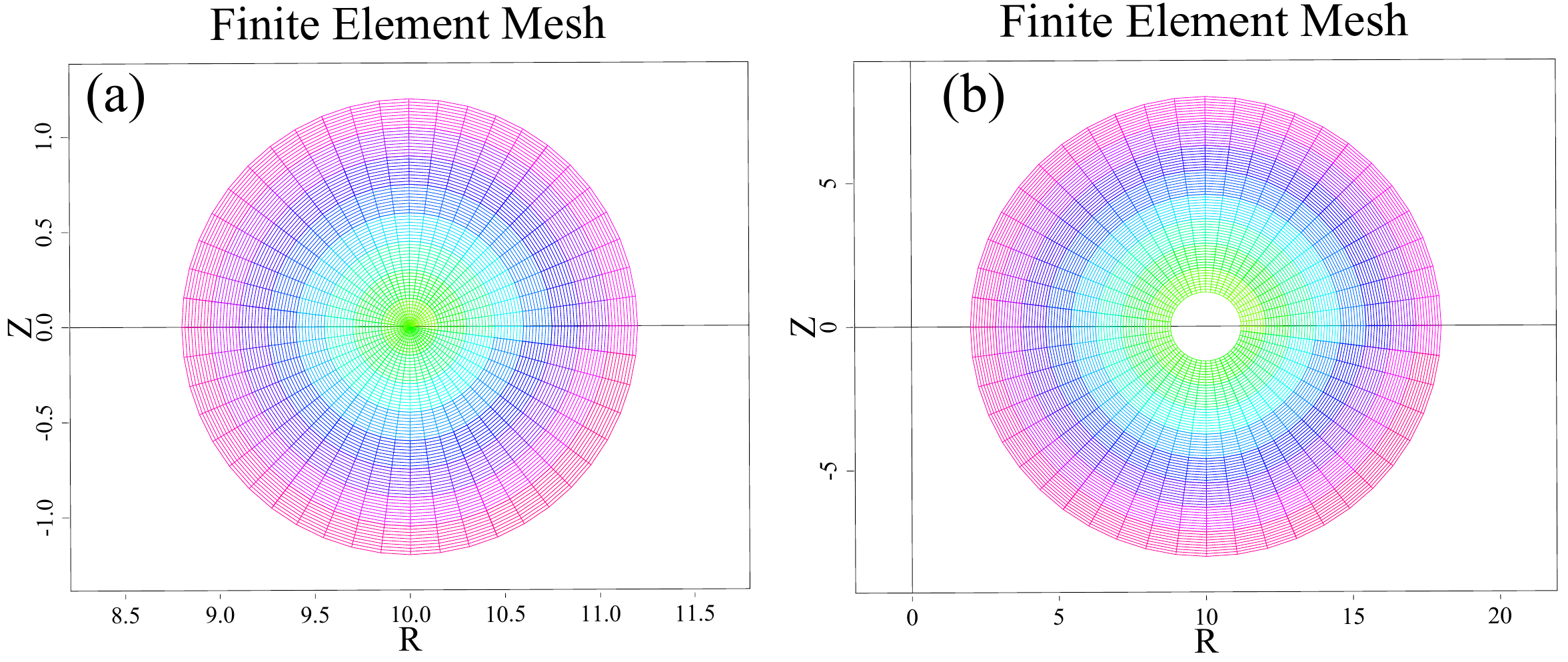}
		\caption{Finite element meshes for (a) the plasma and the inner vacuum regions, and (b) the external vacuum region used in NIMROD calculations. The axial unit is meter.}
		\label{fig:grid}
	\end{figure}

The circular cross section of the cylindrical configuration in the $\left( r, \theta\right) $ plane is discretized using a two-dimensional mesh of $72\times 48$ bi-quintic finite elements (figure~\ref{fig:grid}), and fourier expansion is employed in the axial direction. Thus for each axial or toroidal mode number $n$ set in the NIMROD calculations next, no particular azimuthal or poloidal mode number $m$ is specified in the initial perburbation on the two-dimensional finite element mesh.
		
	\section{Comparisons between theory and NIMROD calculations}\label{sec:5}

\begin{figure}[h]
		\centering
		\includegraphics[width=1\linewidth]{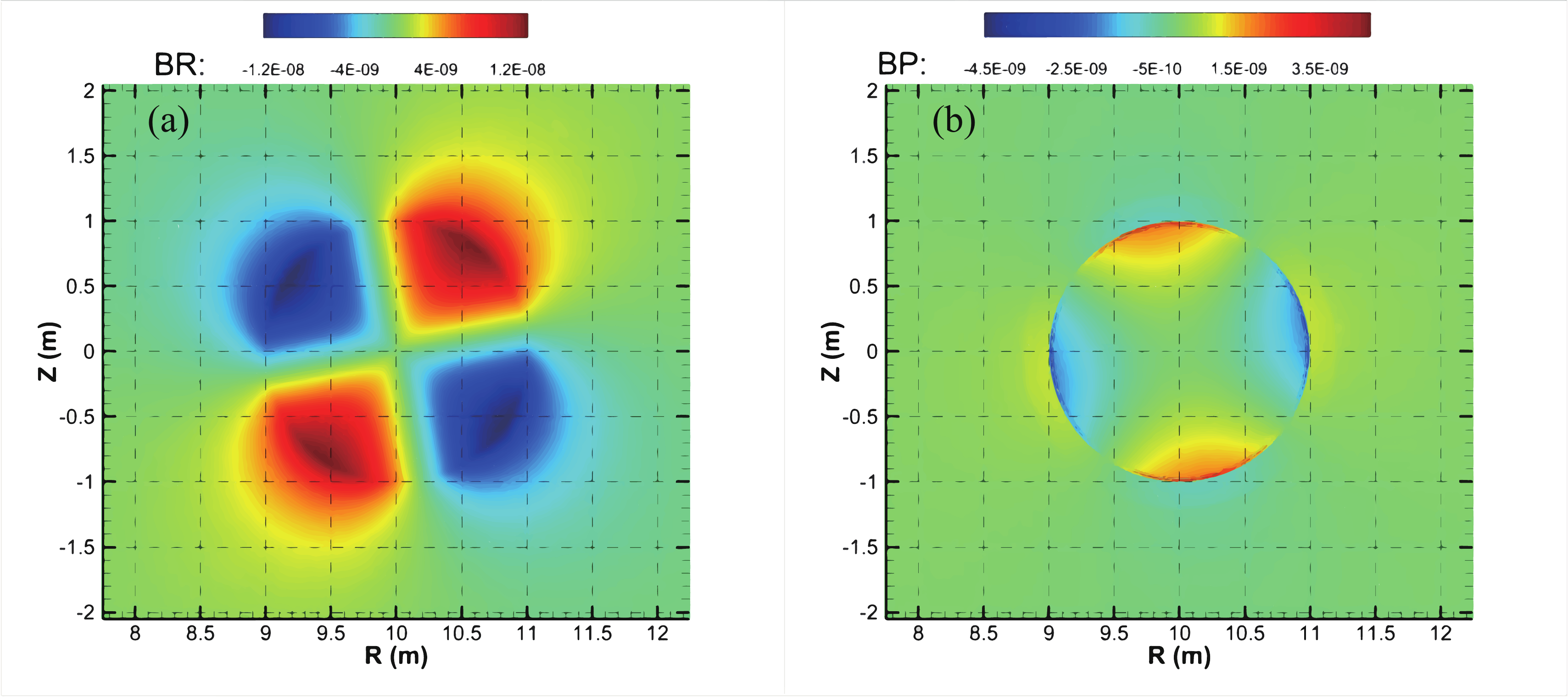}
		\caption{Color contours of (a) the radial component of perturbed $B_r$ and (b) the toroidal component of
			perturbed $B_{\phi}$ for the $n = 1$ resistive wall mode in cylindrical geometry from a NIMROD calculation for the equilibrium $q$ profile using equation~(\ref{eq:dxs representation}) with $q_0=q_a=1.1$, $q_{\rm min}=1.05$, $r_{\rm min}=1.05$ in presence of a resistive wall at the location $r_w/a=1.2$.}
		\label{fig:br-bphi}
	\end{figure}

\begin{figure}[h]
		\centering
		\includegraphics[width=1\linewidth]{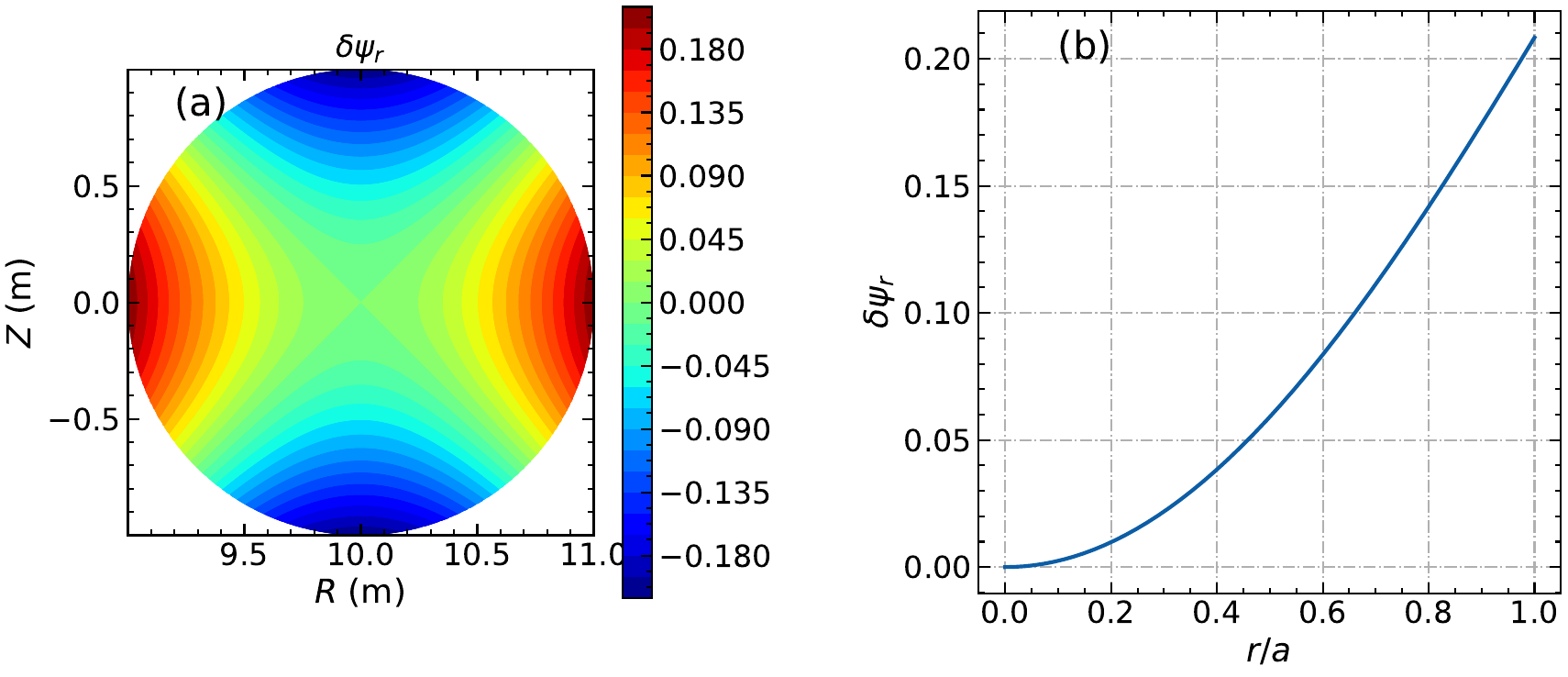}
		\caption{The (a) contour and (b) profile of the perturbed magnetic flux function of a 2/1 RWM obtained from theory using equation~(\ref{eq:RWM-pert-dxs}).}
		\label{fig:theory-eig}
	\end{figure}

	We employ Eqs.~(\ref{eq:rwm-dr}) and (\ref{eq:RWM-DR-dxs}) along with the NIMROD code to compute and compare the effects of magnetic shear on the $n=1$ RWMs, where the inner resistive wall location $ r_w/a=1.2 $ and the wall resistivity parameter $ v_{\rm wall}=500 $~m/s. For $q_{ a} < 2$, the magnetic perturbation contours of both its normal and tangential components to the magnetic flux surface show that the resistive wall radially connects the $m=2$ mode structures located in both inner and external vacuum regions (figure~\ref{fig:br-bphi}), which is consistent with the two-dimensional and one-dimensional distributions of the perturbed magnetic flux function computed from theory using equation~(\ref{eq:RWM-pert-dxs}) (figure~\ref{fig:theory-eig}).
	
	\subsection{Effect of reversed magnetic shear in the core region}\label{sec:5.1}
	
\begin{figure}
		\centering
		\includegraphics[width=0.7\linewidth]{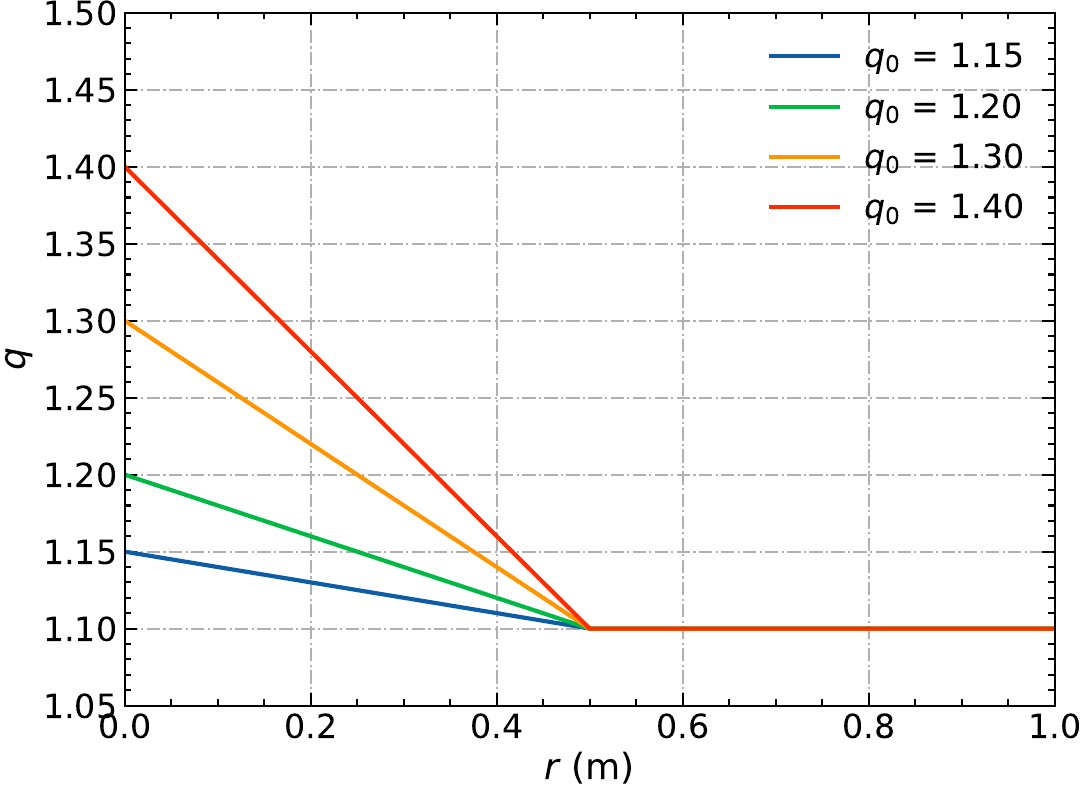}
		\caption{Profiles of the safety factor as the function of minor radius in equation~(\ref{eq:zhexian representation}) for various safety factor values at magnetic axis.}
		\label{fig:function-q01}
	\end{figure}

\begin{figure}
		\centering
		\includegraphics[width=0.7\linewidth]{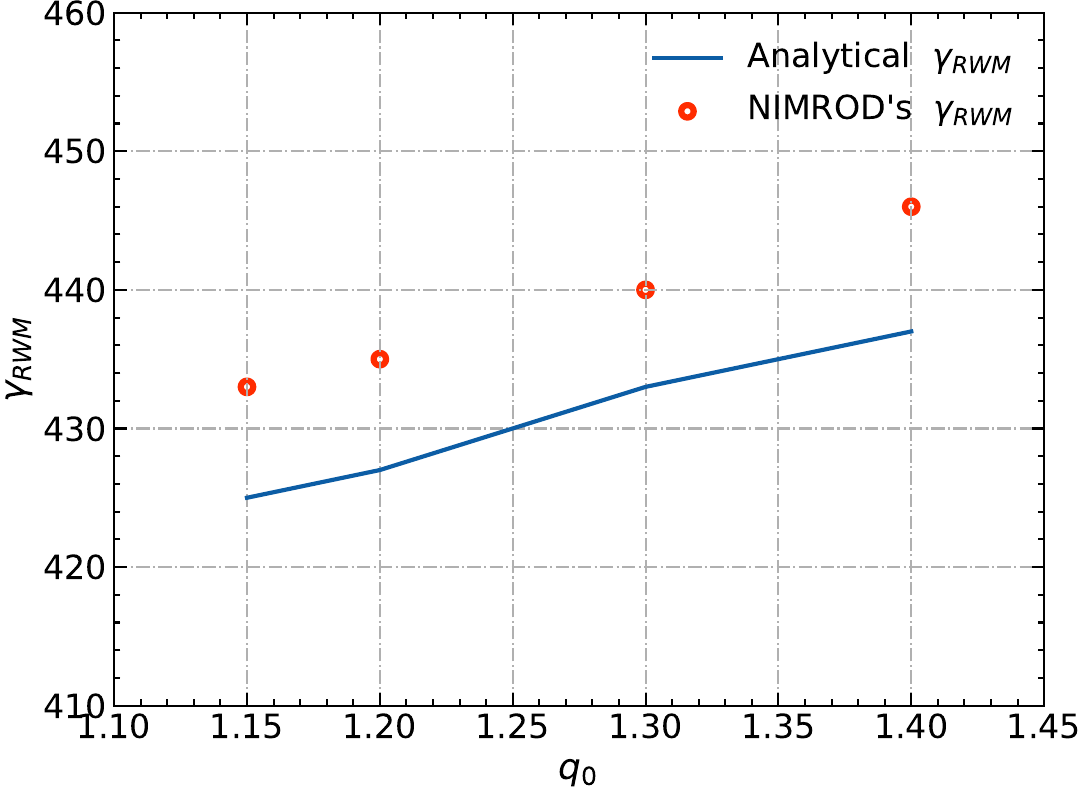}
		\caption{RWM growth rates as functions of $ q_0 $ obtained from the NIMROD calculations (red circles) and the analytical solutions (blue solid line) for the $q$ profiles in figure~\ref{fig:function-q01}.}
		\label{fig:change-q01}
	\end{figure}

\begin{figure}
		\centering
		\includegraphics[width=0.7\linewidth]{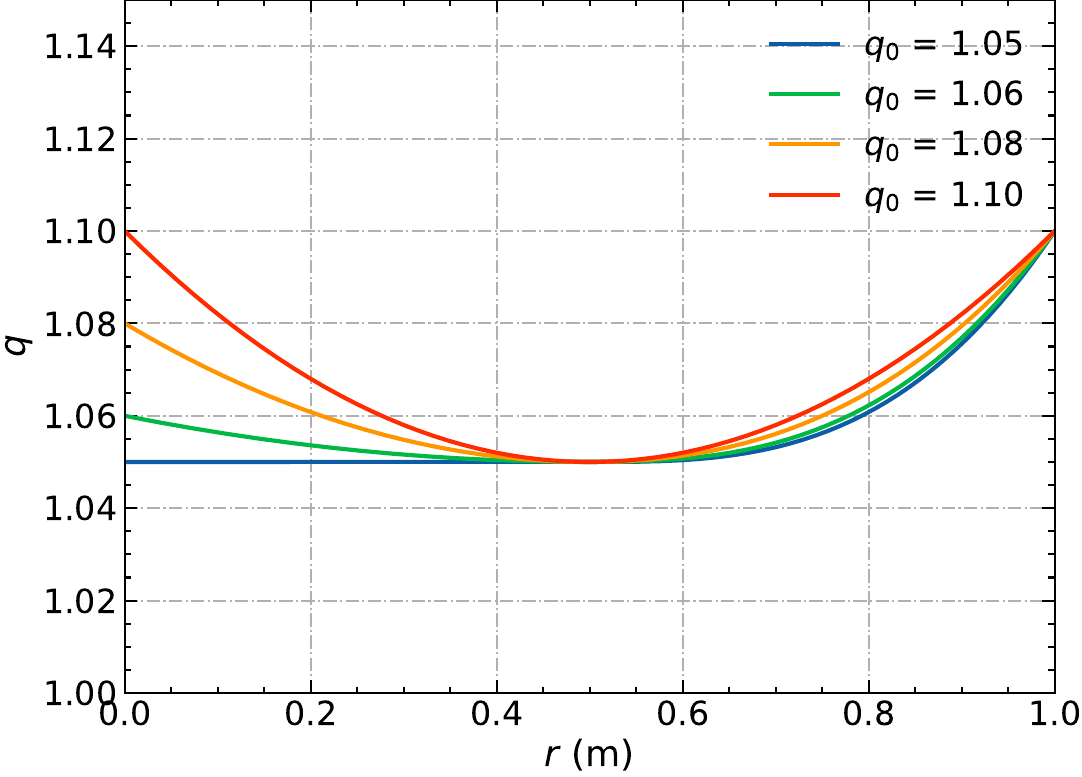}
		\caption{Profiles of the safety factor as the function of minor radius in equation~(\ref{eq:dxs representation}) for various safety factor values at magnetic axis.}
		\label{fig:function-q-smooth}
	\end{figure}

\begin{figure}
		\centering
		\includegraphics[width=0.7\linewidth]{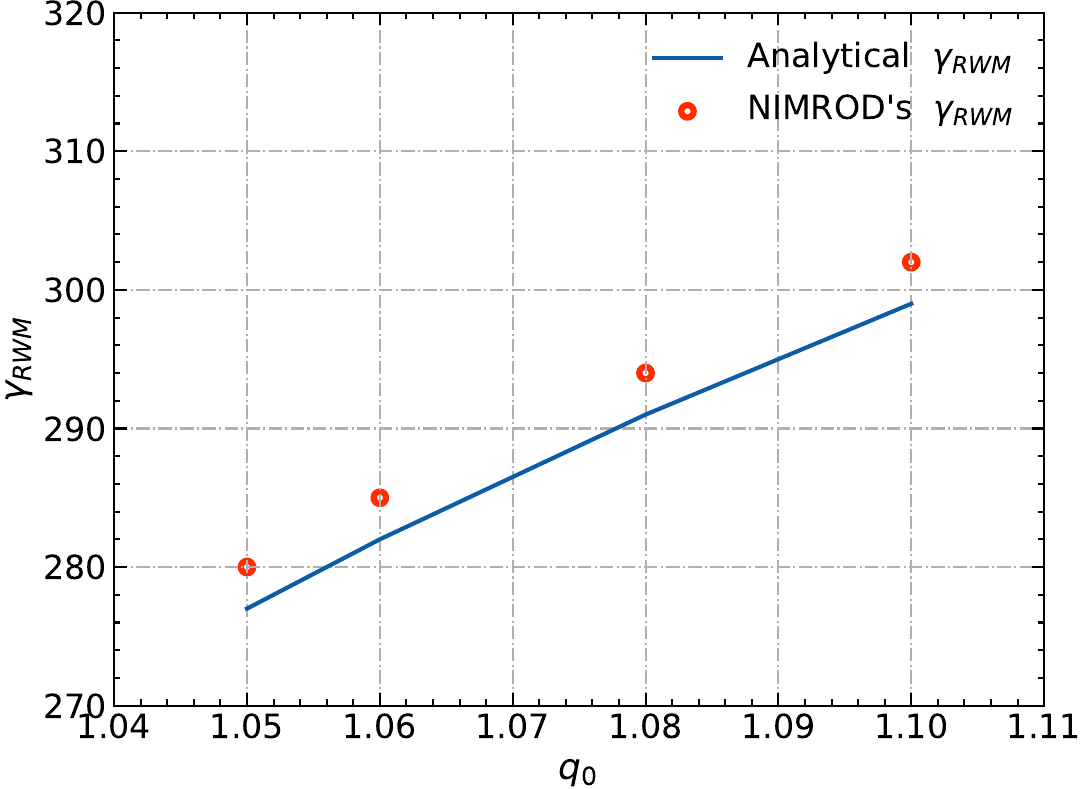}
		\caption{RWM growth rates as functions of $ q_0 $ obtained from the NIMROD calculations (red circles) and the analytical solutions (blue solid line) for the $q$ profiles in figure~\ref{fig:function-q-smooth}.}
		\label{fig:change-q-smooth}
	\end{figure}

\begin{figure}
		\centering
		\includegraphics[width=0.7\linewidth]{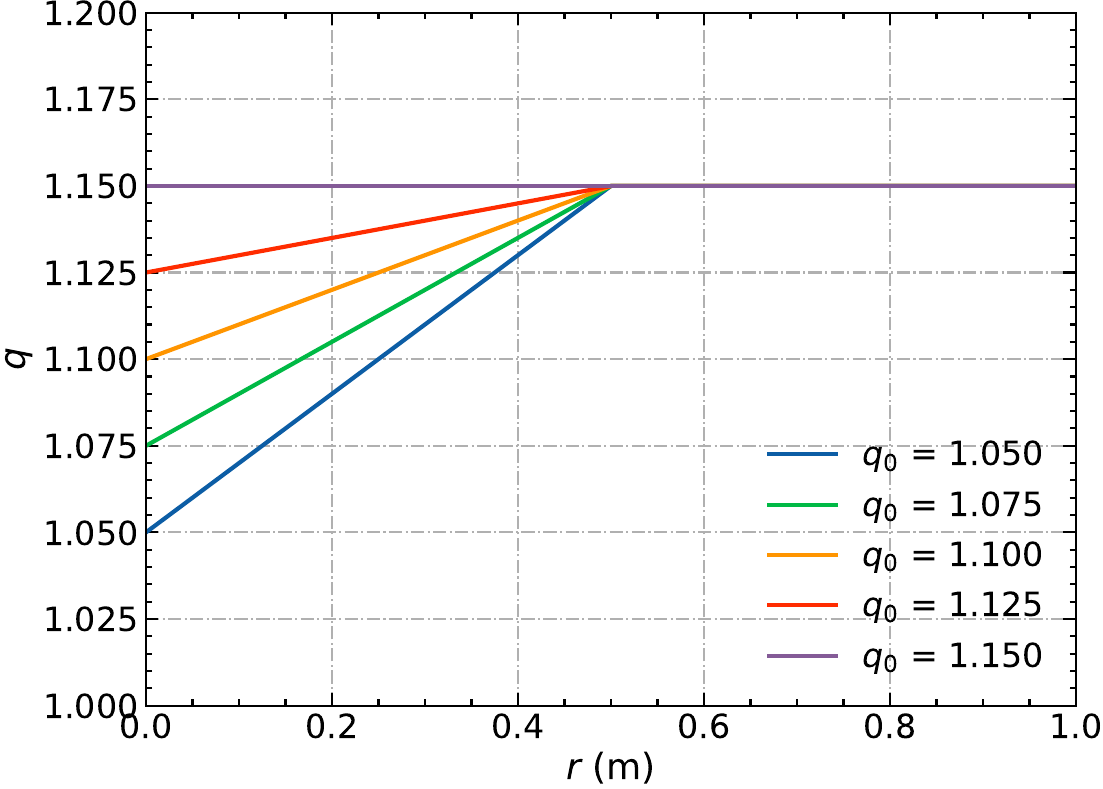}
		\caption{Profiles of the safety factor as the function of minor radius in equation~(\ref{eq:zhexian representation}) for various safety factor values at magnetic axis.}
		\label{fig:function-q02}
	\end{figure}

\begin{figure}
		\centering
		\includegraphics[width=0.7\linewidth]{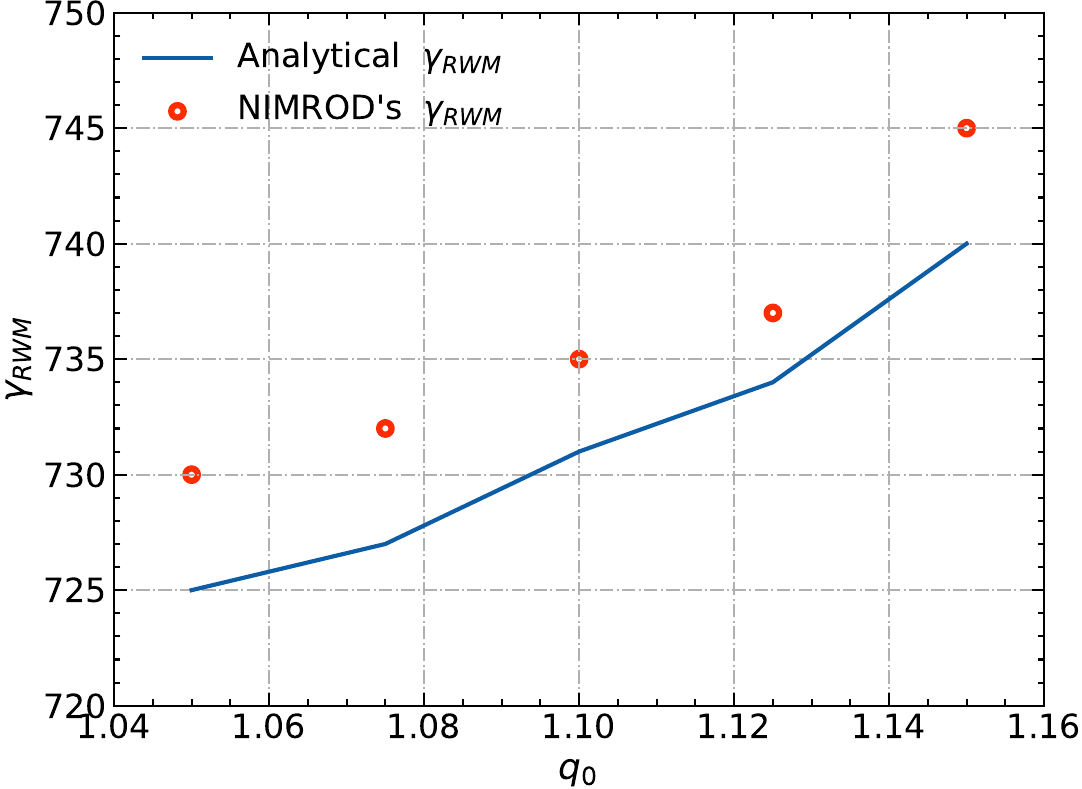}
		\caption{RWM growth rates as functions of $ q_0 $ obtained from the NIMROD calculations (red circles) and the analytical solutions (blue solid line) for the $q$ profiles in figure~\ref{fig:function-q02}.}
		\label{fig:change-q02}
	\end{figure}

	To analyze the influence of magnetic shear in the core region of plasma on RWM stability, we employ the $q$-profiles in equation~(\ref{eq:zhexian representation}) (figure~\ref{fig:function-q01}) and equation~(\ref{eq:dxs representation}) (figure~\ref{fig:function-q-smooth}). For the same safety factor value at plasma edge $ q_f $ and the radial range of reversed magnetic shear $ d $, the safety factor in plasma core $ q_0 $ is varied. The computed results, as depicted in figures~\ref{fig:change-q01} and \ref{fig:change-q-smooth}, exhibit consistency between the theory and NIMROD calculations. For an enhanced reversed magnetic shear in the core, the RWM growth rate increases weakly. In the case of positive magnetic shear configurations (figure~\ref{fig:function-q02}), positive magnetic shear often exhibit a stabilizing effect (figure~\ref{fig:change-q02}), in contrast to the reversed magnetic shear.
	\subsection{Effect of zero magnetic shear at plasma edge}\label{sec:5.2}

\begin{figure}
		\centering
		\includegraphics[width=0.7\linewidth]{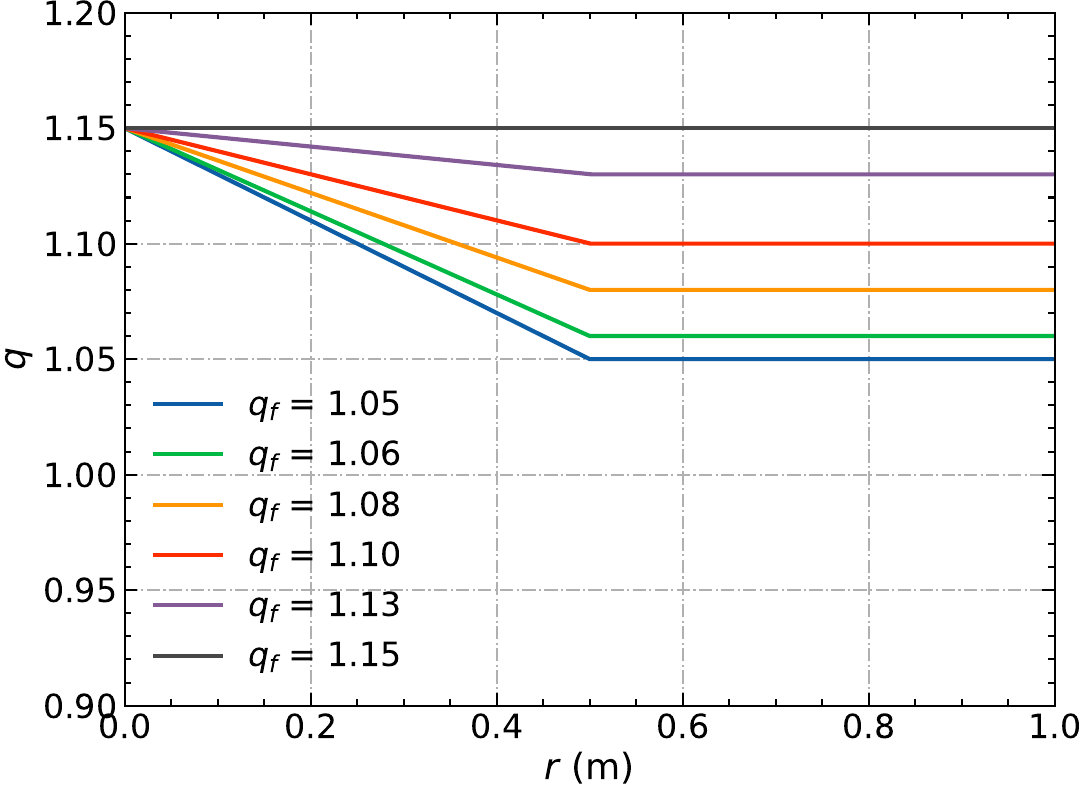}
		\caption{Profiles of the safety factor as the function of minor radius in equation~(\ref{eq:zhexian representation}) for various safety factor values at the edge flat-$q$ region.}
		\label{fig:function-qf}
	\end{figure}

\begin{figure}
		\centering
		\includegraphics[width=0.7\linewidth]{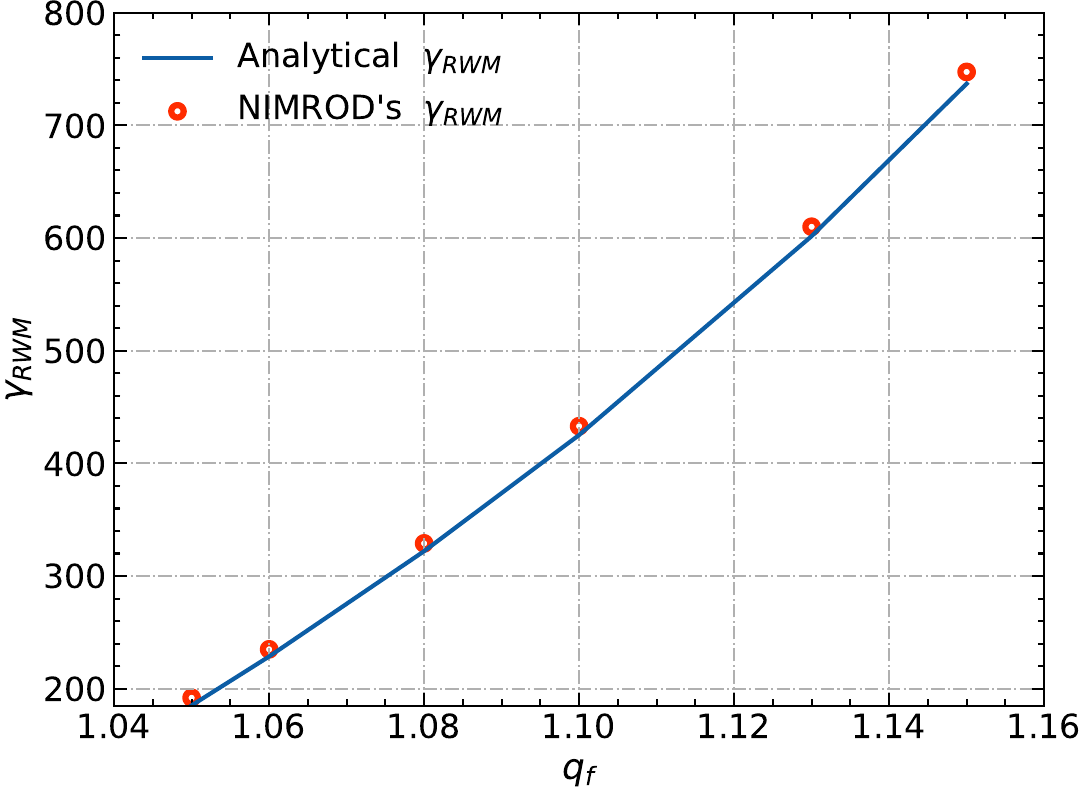}
		\caption{RWM growth rates as functions of $ q_f $ obtained from the NIMROD calculations (red circles) and the analytical solutions (blue solid line) for the $q$ profiles in figure~\ref{fig:function-qf}.}
		\label{fig:change-qf}
	\end{figure}

\begin{figure}
		\centering
		\includegraphics[width=0.7\linewidth]{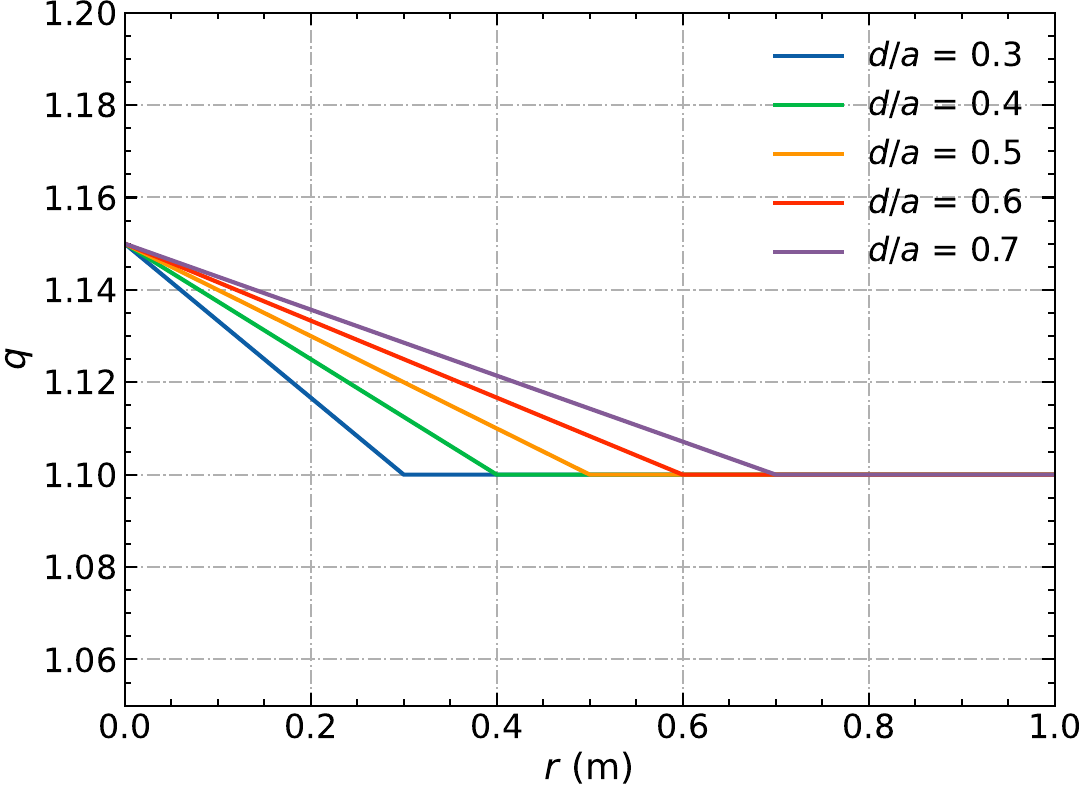}
		\caption{Profiles of the safety factor as the function of minor radius in equation~(\ref{eq:zhexian representation}) for various width of the reversed magnetic shear or the edge flat-$q$ region.}
		\label{fig:function-d-un}
	\end{figure}

\begin{figure}
		\centering
		\includegraphics[width=0.7\linewidth]{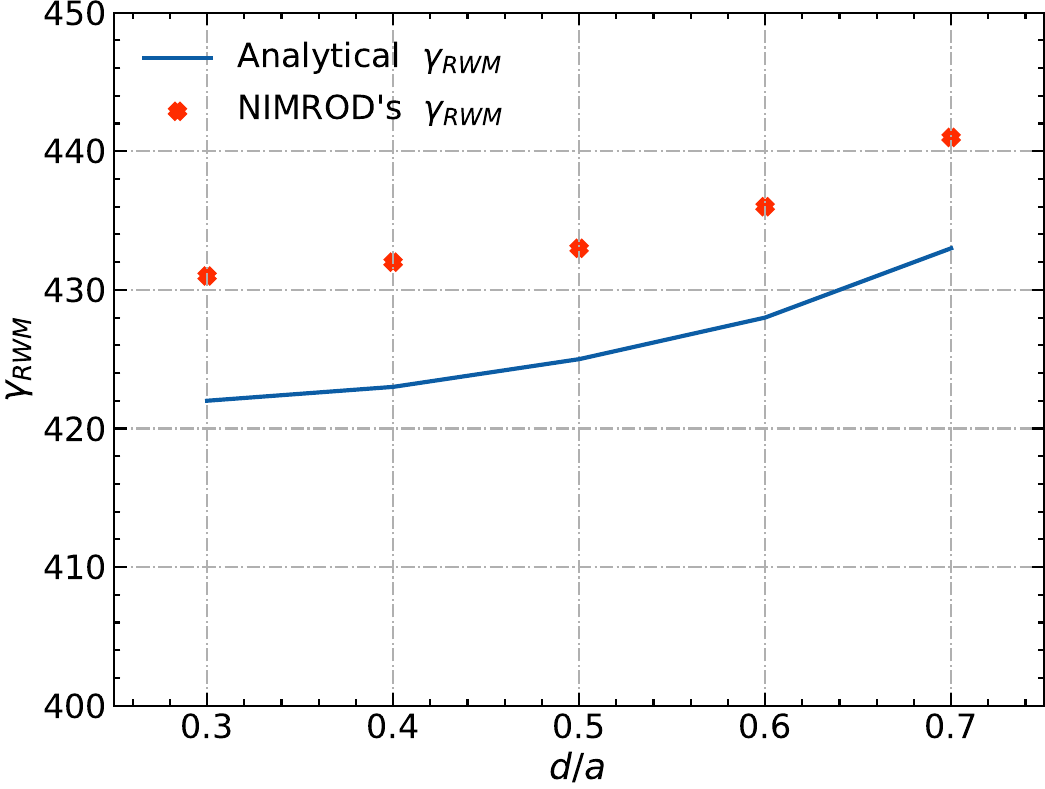}
		\caption{RWM growth rates as functions of $d$ obtained from the NIMROD calculations (red circles) and the analytical solutions (blue solid line) for the $q$ profiles in figure~\ref{fig:function-d-un}.}
		\label{fig:change-d-un}
	\end{figure}

	The impact of the flat-$q$ region at plasma edge on RWM stability is analyzed using the first safety factor profile in equation~(\ref{eq:zhexian representation}). The safety factor at magnetic axis, denoted as $ q_0 $, and the radial range of the reversed magnetic shear region $ d $ are kept fixed, while the value of the boundary flat-$q$, denoted as $ q_f $, is varied. As demonstrated in section~\ref{sec:5.1}, RWM becomes more unstable in presence of enhanced reversed magnetic shear in the plasma core. However, despite the increased reversal of magnetic shear in the core region as $q_{ f}$ decreases, there is a significant reduction in the growth rates obtained from both analytical theory and NIMROD results which are in good agreement (figures.~\ref{fig:function-qf}-\ref{fig:change-qf}). This implies that lowering the boundary flat-$q$ value has a dominant stabilizing influence on the RWM. The range of the boundary flat-$q$ region measured by $ \left( 1-d/a\right) $, can be also varied while keeping $q_0$ and $q_{ f}$ fixed. It is evident that a widened region of zero shear at the plasma edge contributes to a dominant stabilizing effect that is able to overcome the destabilizing effect of the enhanced reversal of magnetic shear at core region (figures.~\ref{fig:function-d-un}-\ref{fig:change-d-un}).

	\subsection{Effect of positive magnetic shear at plasma edge}

\begin{figure}
		\centering
		\includegraphics[width=0.7\linewidth]{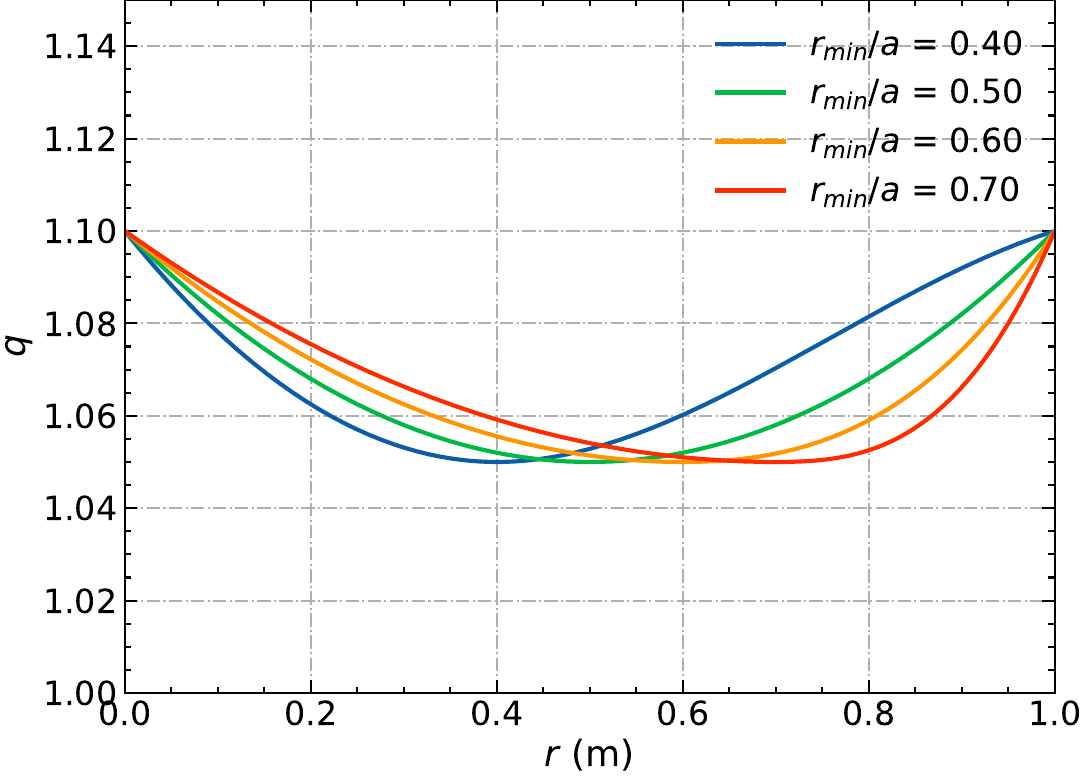}
		\caption{Profiles of the safety factor as the function of minor radius in equation~(\ref{eq:dxs representation}) for various values of $ r_{\rm min} $.}
		\label{fig:function-d-sm}
	\end{figure}

\begin{figure}
		\centering
		\includegraphics[width=0.7\linewidth]{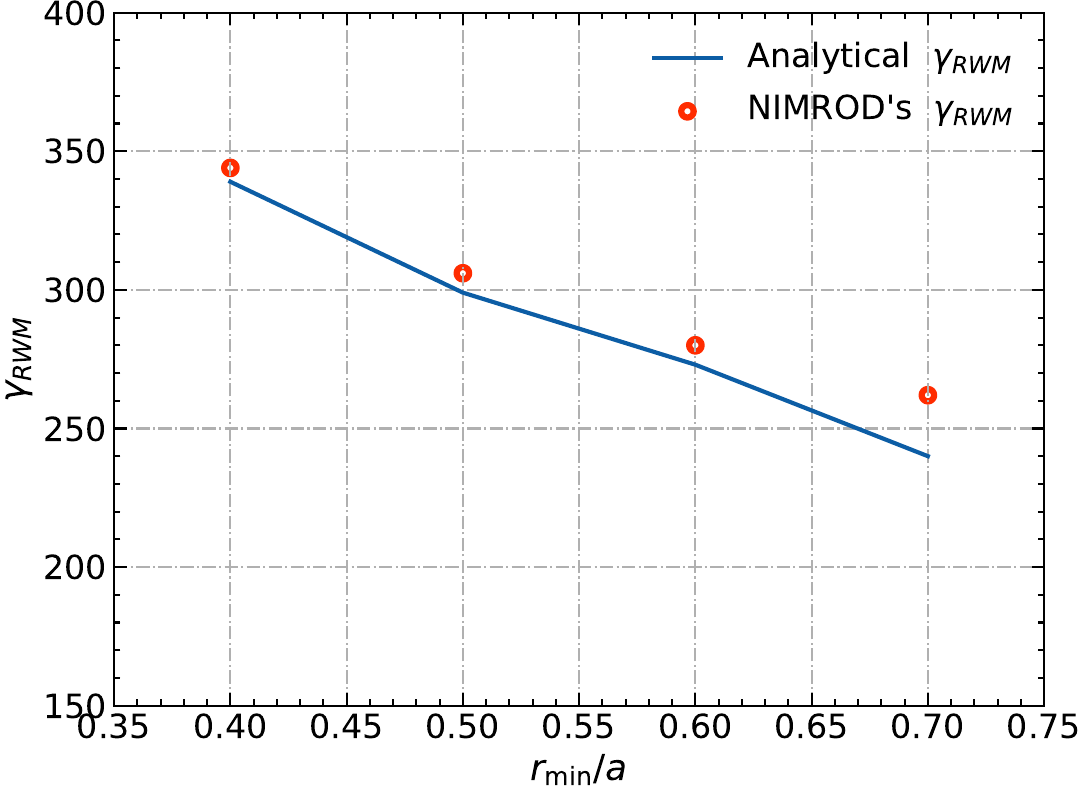}
		\caption{RWM growth rates as functions of $r_{\rm min}$ obtained from the NIMROD calculations (red circles) and the analytical solutions (blue solid line) for the $q$ profiles in figure~\ref{fig:function-d-sm}.}
		\label{fig:change-d-sm}
	\end{figure}

	Utilizing the safety factor profile as given by equation~(\ref{eq:dxs representation}), and keeping the safety factor values at magnetic axis and boundary, as well as the minimum location fixed, here we vary only the radius of the minimum location $r_{\rm min}$ (figure~\ref{fig:function-d-sm}) to investigate the impact of the magnetic shear near plasma boundary on RWM. In contrast to section~\ref{sec:5.2} where the growth rate of RWM gradually increases with the widening in $d$ or $r_{\rm min}$ (figure~\ref{fig:change-d-un}), the results here reveal a gradual decrease in RWM's growth rate with the presence of increasing positive magnetic shear at the plasma boundary (figures.~\ref{fig:function-d-sm}-\ref{fig:change-d-sm}). This observation suggests that the positive magnetic shear at the boundary often has a more dominant stabilizing effect.
	
\begin{figure}
		\centering
		\includegraphics[width=0.7\linewidth]{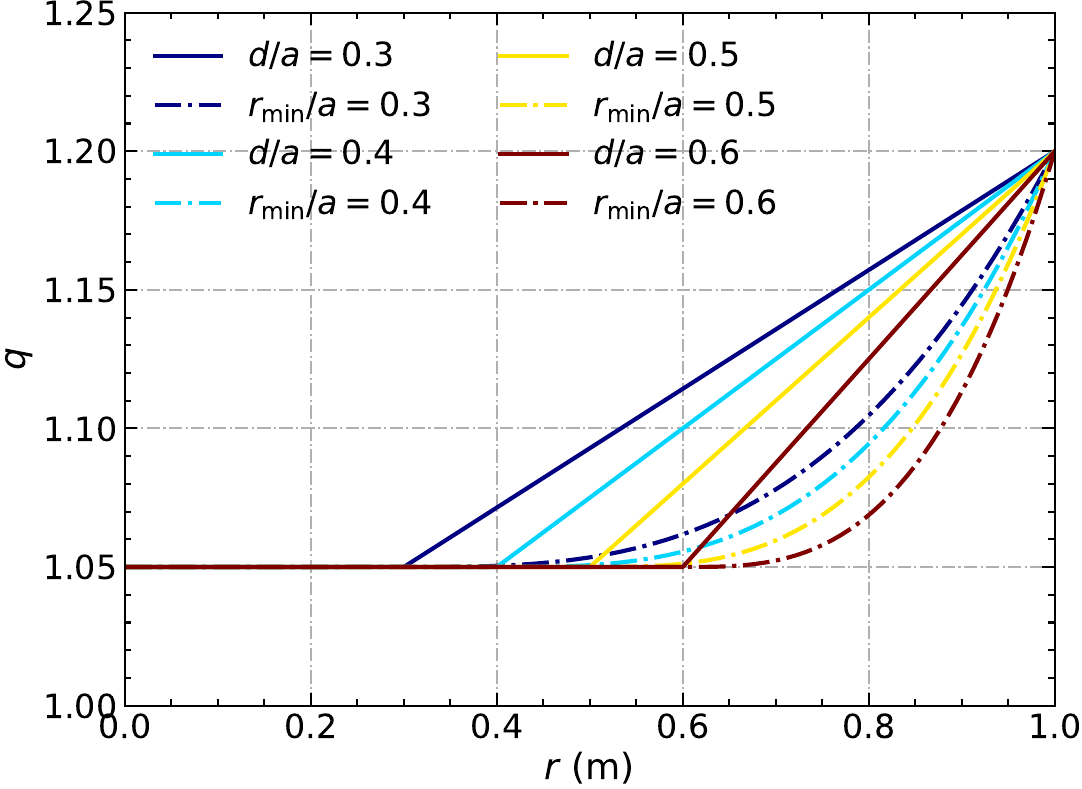}
		\caption{Profiles of the safety factor as the function of minor radius in equation~(\ref{eq:zhexian representation}) (solid line) and equation~(\ref{eq:dxs representation}) (dotdash line) for various values of $d/a$ or $r_{\rm min}/a$ from 0.3 to 0.6.}
		\label{fig:functionrminvariation}
	\end{figure}

\begin{figure}
		\centering
		\includegraphics[width=0.7\linewidth]{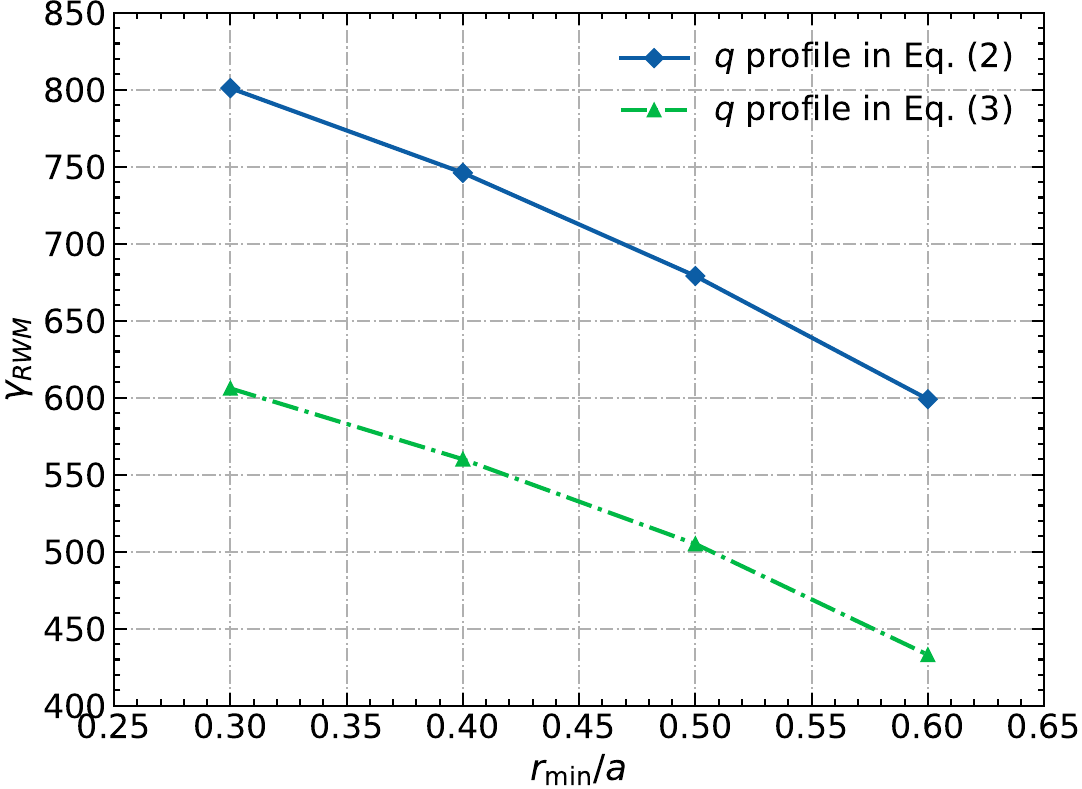}
		\caption{RWM growth rates as functions of $d/a$ or $ r_{\rm min}/a $ for the various $ q $ profiles in equation~(\ref{eq:zhexian representation}) (blue solid line) and equation~(\ref{eq:dxs representation}) (green dotdash line).}
		\label{fig:smooth}
	\end{figure}

	To further examine above finding, several $q$ profiles in presence of varying edge positive shear only are compared (figure~\ref{fig:functionrminvariation}). The results clearly demonstrate that the growth rate of RWM decreases gradually with increasing positive shear near edge (figure~\ref{fig:smooth}), thus confirming its stabilizing effect.

	\section{Conclusions and discussion}\label{sec:6}

	In summary, the analytical dispersion relation of the current-driven resistive wall mode is derived in a cylindrical geometry and compared with the NIMROD calculations with good agreement. The enhanced reversed magnetic shear in the core and a narrower region of zero shear at the edge are found to be the most destabilizing on the RWM. In contrast, the positive magnetic shear near the plasma boundary is found stabilizing on the mode.
	
	The analytical results in this work are mainly applicable to equilibra with relatively weak magnetic shear. For more realistic equilibrium configurations, such as those in CFETR and ITER, it is necessary to further extend the analytical results. In addition, we also plan to evaluate the pressure or $\beta$ effects and the influence of the non-vacuum SOL region on RWM.
	
	\section*{Acknowledgements}
	
	We are grateful for the supports from the NIMROD team. This work is supported by the National MCF Energy R\&D Program of China under Grant No.~2019YFE03050004, the National Natural Science Foundation of China Grant No.~51821005, and the U.S. Department of Energy Grant No.~DE-FG02-86ER53218. The computing work in this paper is supported by the Public Service Platform of High Performance Computing by Network and Computing Center of HUST.
	\clearpage
	\section*{References}
	\bibliography{sample-1}
	\clearpage
	\clearpage
	\clearpage
	\clearpage
	\clearpage
	\clearpage
	\clearpage
	\clearpage
	\clearpage
	\clearpage
	\clearpage
	\clearpage
	\clearpage
\end{document}